\documentclass{nature}

\usepackage{amsfonts,amssymb,amsmath,amsthm,amscd}

\usepackage{graphicx}

\newcommand{\Hh}{\mathcal{H}}

\newcommand{\ox}{\otimes}
\newcommand{\bra}[1]{\langle #1\vert}                  
\newcommand{\ket}[1]{\vert #1 \rangle}    
\newcommand{\HR}{\text{HR}}
\newcommand{\M}{\text{M}}
\newcommand{\BH}{\text{BH}}

\DeclareMathOperator{\Tr}{Tr}

\title{Hawking radiation and the quantum marginal problem}

\author{Erik Aurell$^{1}$, Micha\l\ Eckstein${}^{2,3}$, Pawe\l\ Horodecki${}^{4,5}$}

\makeatletter
\let\saved@includegraphics\includegraphics
\AtBeginDocument{\let\includegraphics\saved@includegraphics}
\renewenvironment*{figure}{\@float{figure}}{\end@float}
\makeatother

\begin{document}

\maketitle

\begin{affiliations}
 \item \mbox{KTH -- Royal Institute of Technology, AlbaNova University Center, SE-106 91 Stockholm, Sweden}
 \item Institute of Theoretical Physics, Jagiellonian University, {\L}ojasiewicza 11, 30-348 Krak\'ow, Poland
 \item Copernicus Center for Interdisciplinary Studies, ul. Szczepa\'nska 1/5, 31-011 Krak\'ow, Poland
 \item International  Centre  for  Theory  of  Quantum  Technologies,  University  of  Gda\'nsk,  Wita  Stwosza  63,  80-308  Gda\'nsk,  Poland
 \item Faculty of Applied Physics and Mathematics, National Quantum Information Centre,
Gda\'nsk University of Technology, Gabriela Narutowicza 11/12, 80-233 Gda\'nsk, Poland
\end{affiliations}

\begin{abstract}
In 1974 Steven Hawking showed that black holes emit thermal radiation, which eventually causes them to evaporate.
The problem of the fate of information in this process is known as the ``black hole information paradox''. 
Two main types of resolution postulate either a fundamental loss of information in Nature -- hence the breakdown of quantum mechanics -- or some sort of new physics, e.g. quantum gravity, which guarantee the global preservation of unitarity.
Here we explore the second possibility with the help of recent developments in continuous-variable quantum information.
Concretely, we employ the solution to the Gaussian quantum marginal problem to show that the thermality of all individual Hawking modes is consistent with a global pure state of the radiation. Surprisingly, we find out that the mods of radiation of an astrophysical black hole are thermal until the very last burst. In contrast, the single-mode thermality of Hawking radiation originating from microscopic black holes, expected to evaporate through several quanta, is not excluded, though there are constraints on modes' frequencies. Our result paves the way towards a systematic study of multi-mode correlations in Hawking radiation.
\end{abstract}

\section{Introduction}

When a star runs out of its nuclear fuel the gravitational attraction takes over and the star implodes. If the mass of the stellar remnant exceeds the Tolman--Oppenheimer--Volkoff limit ($\sim$ 2.3 solar masses\cite{TOV}), the gravitational collapse will result in a black hole. According to general relativity, the latter consist of a central spacetime singularity hidden under a trapping horizon -- a surface impenetrable from the inside by any particles or fields\cite{PenroseCollapse}. The area of the horizon continues to grow as the black hole engulfs the surrounding matter. The collapse progresses indefinitely, trapping all infallen information carried by the matter fields in the black hole region.

However, once an event horizon forms, the quantum field theory predicts a steady flux of thermal radiation emitted from the black hole\cite{Hawking74,Hawking75,Wald75}. Its origin is attributed to the  fluctuations of quantum vacuum near the event horizon. These result in spontaneous pair creation, with one of the partners falling into the black hole and the other escaping to infinity. An alternative description of the mechanism involves tunnelling of quantum particles through the event horizon\cite{ParikhWilczek2000}. In consequence, a black hole can be (approximately\cite{Visser15}) seen as a black body. The radiation consists mostly of photons (and, possibly, other free massless elementary particles), as emission of particles of mass $m$ can only happen when the black body temperature exceeds the threshold $m c^2 / k_{\text{B}}$ for their production\cite{Page76}.

The laws of black hole thermodynamics\cite{Bekenstein1973} allow to explicitly compute the temperature of the Hawking radiation\cite{Page05}. For a Schwarzschild black hole $T_{\text{BH}} = \frac{\hbar c^3}{8 \pi k_\text{B} G M}$ depends only upon the black hole's mass $M$ and entails the Planck constant $\hbar$, the gravitational constant $G$, Boltzmann constant $k_\text{B}$ and the speed of light in vacuum $c$. Even for a small black hole of several solar masses the value of $T_{\text{BH}}$ is of the order of $10^{-8} K$, much below the temperature of the Cosmic Microwave Background, $\sim 2.7 K$. In consequence, a newly formed black hole will be acquiring much more energy than it radiates out.
 
However, according to the standard cosmological model, as the Universe expands the matter will get more and more sparse and the average temperature will drop. 
Eventually, the Universe will enter a ``dark epoch'' when the Hawking radiation dominates. A black hole will gradually loose its mass and, finally, evaporate completely. This process is extremely slow --- for stellar-mass black holes it takes roughly $10^{67}$ years --- but becomes more dynamical towards the end since $T_\text{BH}$ grows as the mass decreases. A black hole is thus expected to end its life in a very bright explosion\cite{Hawking74}.

It is also possible that primordial black holes were formed from density fluctuations at a very early stage of the Universe's evolution\cite{Hawking71,CarrHawking1974,Sasaki2018}. Such hypothetical objects could have, depending on the model, masses ranging from the Planck mass $M_\text{P} \approx 2 \cdot 10^{-8}$ kg to several solar masses. Those with an initial mass larger than $10^{11}$ kg would still be present in the Universe, while the ones of a smaller initial mass would have already evaporated.

Furthermore, some theoretical scenarios involving spacetime foam or extra dimensions imply the production of microscopic black holes at the Planck scale\cite{BH_virtual,BH_LHC1,BH_LHC2}. Such hypothetical objects could lead to observable effects in particle colliders\cite{BH_CMS,BH_ATLAS} and future neutrino telescopes\cite{Micro_BH_nu}. Yet another class of Planck scale theories posit the existence of microscopic black holes basing on the physical limits of measurements\cite{DFR94,GUP_BH,Minimal_Length_Review}.

\section{\label{sec:unitarity}The global unitary evolution}

The description of the black hole's evolution via a global unitary transformation (an $S$-matrix) was conceived already in 1980 by Don Page\cite{Page80,Page93}. It implies that quantum information gradually escapes the black hole region and remains concealed in the quantum correlations between the modes of the Hawking radiation.

\pagebreak

Here we follow the same line of reasoning, pinpointing the trouble spot concerning the admissible dynamics of quantum correlations. 
To this end, we adopt a `model-independent' methodology, to some extent inspired by the device-independent approach to quantum information (cf. e.g. Refs.\cite{Bell_Nonlocal,PR_box}). It consists in regarding physical systems as `black boxes' and studying their information-processing properties, without employing any specific physical model. Such an approach allows one to establish powerful information-theoretic principles, which sets universal limitations on the admissible physical models (see e.g. Ref.\cite{InformationCausality}). In this spirit, we regard a black hole as a \emph{quantum}-information processing device and seek to unveil general features, implied by the assumptions of global unitarity and single-mode thermality, which would constrain the possible black hole evaporation scenarios.


In quantum mechanics an isolated system is modelled by a pure state, i.e. a vector in a Hilbert space. It describes the full knowledge about the system in terms of quantum information.  Let us consider, from the perspective of a distant observer, the isolated system initially consisting of a collapsing star accompanied with all matter which will eventually fall into the forming black hole. The relevant Hilbert space is decomposed into three segments $\Hh = \Hh_{\text{M}} \ox \Hh_{\text{BH}} \ox \Hh_{\text{HR}}$ tailored for the infalling matter outside of the trapping horizon, the quantum systems associated with the black hole region and the Hawking radiation, respectively. We take the space $\Hh_{\text{HR}}$ to be infinite-dimensional, as usual in the considerations involving quantum thermal radiation\cite{GaussQI}. The BH sector includes the quantum degrees of freedom of the particles under the trapping horizon, as well as possibly other quantum registers associated e.g. with the quantum gravity --- see e.g. Ref.\cite{FOP2021} and references therein for a more detailed discussion.

Initially, the global pure state of the system is of a product form
\begin{align*}
\ket{\Psi_\text{in}} = \ket{\psi_0}_\M \ox \ket{0}_\BH \ox \ket{0}_\HR,
\end{align*}
where $\ket{0}$ denotes the vacuum. Once the trapping horizon begins to form the quantum information starts flowing from the M sector to the BH sector. As the quantum systems constituting the infalling matter are generically entangled, the resulting global state will involve entanglement between M and BH,
\begin{align*}
\ket{\Psi^{(0)}} = \sum_i a_i \, \ket{\psi_i^{(0)}}_\M \ox \ket{\chi_i^{(0)}}_\BH \ox \ket{0}_\HR.
\end{align*}
When a quantum of Hawking radiation is emitted at the event horizon, it is entangled with its partner concealed in the black hole region. The pair forms a pure state $\sum_j b_j \ket{\eta_j}_\BH \ox \ket{\phi_j}_\HR$. Shortly afterwards, the infalling particles interact with the matter in the black hole region and the outgoing Hawking radiation becomes entangled with other quantum systems in BH. The total state will then have the form
\begin{align*}
\ket{\Psi^{(1)}} = \sum_{i,j} c_{ij} \, \ket{\psi_i^{(1)}}_\M \ox \ket{\chi_i^{(1)}}_\BH \ox \ket{\phi_j^{(1)}}_\HR,
\end{align*}
known as ``entangled entanglement''\cite{EntEnt,EntEntExp}.

As the subsequent quanta of radiation emitted at the horizon, the quantum information gets gradually transferred to the Hawking radiation. Eventually, all of the matter from $\M$ will fall in the black hole and the latter will evaporate through the emission of a finite number $N$ of radiation quanta. The final quantum state of the process will then be of the form
\begin{align*}
\ket{\Psi_\text{out}} = \ket{0}_\M \ox \ket{0}_\BH \ox \ket{\Phi}_\HR, 
\end{align*}
with some complex pure state of Hawking radiation exhibiting multi-mode entanglement,
\begin{align}\label{final}
\ket{\Phi} = \!\!\!\sum_{k_1, k_2, \ldots, k_N}\!\!\!f_{k_1, k_2, \ldots, k_N} \ket{\phi_{k_1}^{1}} \ox \ket{\phi_{k_2}^{2}}  \ox \cdots \ox \ket{\phi_{k_N}^{N}}.
\end{align}
\indent The final state \eqref{final} contains all quantum information initially encoded in the infalling matter and subsequently hidden under the trapping horizon. Recent computations\cite{Padmanabhan} based on the semi-classical approach indicate how some information about the initial state of the collapse can be obtained through studying the distortions of the black hole radiation. It is well-understood, however, that a complete retrieval of the encoded quantum information would require a fully quantum treatment of the black hole.

At intermediate stages of the process, the Hawking radiation remains entangled with the black hole sector, as well as with the infalling matter if the collapse is ongoing. Consequently, during the evaporation phase the quantum information about the initial state of the collapse is concealed in a composite system involving the black hole sector, the outgoing radiation \emph{and} correlations between them. It is generally expected\cite{Page93,Maldacena20} that the `information flow' becomes significant only after the Page time, which marks roughly the half of the black hole's lifetime. It was also suggested that the information decoding might actually happen only towards the very end of the evaporation process\cite{Karol_BH}.

The information-theoretic analysis based solely on the assumption of global unitarity does not provide any hint about the actual dynamics of (quantum) information. Both models involving a slow gradual transfer of information and those predicting the onset of the information flow are, in principle, conceivable. 

On the other hand, whereas the global evolution $\ket{\Psi_\text{in}} \mapsto \ket{\Psi_\text{out}}$ is --- by assumption --- unitary, 
it cannot be described within the standard LOCC (Local Operations and Classical Communi-cation\cite{Horodecki}) paradigm for (quantum) information processing. Indeed, the Hawking modes are emitted independently, hence they are initially in a product state. They cannot get entangled through any known local quantum process involving the BH and M sectors, because the corresponding unitary transformation would have the form $U_{\text{M,BH}} \ox \text{id}_{\text{HR}}$. The direct interaction between the outgoing Hawking modes, such as light-by-light scattering, which could lead to entanglement, is negligible. Similarly, the interaction between Hawking radiation and the infalling matter (sector M), which leads to grey-body corrections\cite{Visser15}, is not strong enough to restore the global unitarity\cite{Mathur_2009}. In conclusion, the mechanism behind the global unitary evolution necessarily involves a departure from \emph{local} quantum physics, and hence quantum field theory (cf. Ref.\cite{Harlow}).

Finally, let us point out that a description of the BH sector consistent with the Bekenstein--Hawking entropy requires the activation of a huge number, $\mathcal{N} \sim \exp( 4\pi M^2/M_P^2)$, of (quantum) degrees of freedom during the collapse stage. The super-exponential growth of entropy cannot be described in terms of the known matter\cite{tHooft93}, which strongly suggests that some model of quantum gravity is eventually needed to account for it\cite{Maldacena20}. An information-theoretic model explaining the BH entropy in terms of the entanglement entropy between the matter and gravity sectors was presented in our previous work\cite{FOP2021}.

We have explored the general consequences implied by the assumption that the quantum information is preserved during the collapse. We now turn towards its compatibility with the mode-by-mode thermality of the Hawking radiation.  

\section{Thermality as the quantum marginal problem}

Suppose that the Hawking radiation does end up in a global pure state of the form \eqref{final}. How different is this state from a thermal state predicted\cite{Wald75} by the quantum field theory? The general expectation (cf. Refs.\cite{Harlow,Raju2020}) is that these states would be almost indistinguishable for an external observer, because the number of modes is very large. Indeed, the total number of particles emitted by a black hole of mass $M$ during the evaporation process can be estimated\cite{Muck16} as $N \sim  \tfrac{5 G}{\hbar c} M^2$. Even for the smallest stellar-mass black holes this number is huge, $N \gtrsim 10^{77}$.

A rigorous argument in this direction was given by Page in Ref.\cite{Page93A} (cf. also Ref.\cite{Harlow}). The Page theorem shows that a random pure state $\ket{\psi_{AB}}$ on a finite-dimensional composite Hilbert space $\Hh_A \otimes \Hh_B$ will have the local reductions $\rho^{A,B} = \Tr_{B,A} \ket{\psi_{AB}} \bra{\psi_{AB}}$ being very close to a totally mixed state, provided that $\dim \Hh_A \ll \dim \Hh_B$.

However, in the complete treatment of quantized radiation one needs to work with infinite dimensional Hilbert spaces, for which this argument does not directly apply. Furthermore, the Hawking temperature rises as the black hole looses mass, so one should also check whether the temperatures associated with the thermal modes are compatible with the Hawking's formula for $T_{\text{BH}}$ in every epoch. In this work, we employ the modern framework of continuous-variables quantum information\cite{ContQI} to overcome these issues.

A state of radiation, pure or mixed, is called \emph{Gaussian} if its corresponding Wigner function is a Gaussian (see Ref.\cite{GaussQI} for the details). A \emph{thermal} state is a Gaussian state, which maximises the von Neumann entropy for a fixed energy. Specifically, a single-mode thermal state can be written in the standard harmonic oscillator basis as
\begin{align}
\rho = \frac{1}{Z} \sum_{n=0}^{\infty} e^{- \tfrac{\hbar \omega}{k_B T} \big(n+\tfrac{1}{2}\big)} \ket{n}\bra{n} = \sum_{n=0}^{\infty} \frac{2 b^n}{\big(b+2 \big)^n} \ket{n}\bra{n}, &&\text{with}&& b = 2 \Big( e^{\tfrac{\hbar \omega}{k_B T}} - 1 \Big)^{-1}.
\end{align}

Now, given a multi-mode state of radiation one can ask when is it compatible with thermal single-mode reductions with given temperatures. This is an instance of the \emph{Quantum Marginal Problem} (know also as the $N$-representability problem),
which was a significant stumbling block in quantum chemistry, until the solution in the framework of quantum information was given by Klyachko\cite{Klyachko2004,Klyachko2006}. For infinite-dimensional systems the quantum marginal problem was posed and solved, for an arbitrary number of bosonic modes, by Eisert et al.\cite{Eisert2008}.

The answer is as follows: Suppose that state $\ket{\Phi}$
is a pure Gaussian state of $N$ modes, then the reduced state $\rho_n$ of every mode is a (mixed) Gaussian state. This is consistent if and only if for $n \in \{1,2,\ldots,N\}$ one has
\begin{align}\label{Eisert}
b_n \leq \sum_{m \neq n} b_m, \quad \text{with} \quad b_n = 2 \left( \exp\!\left(\tfrac{\hbar \omega_n}{k_\text{B} T_n}\right) - 1 \right)^{-1},
\end{align}
where $\omega_n$ is the frequency of the mode and $T_n$ the respective temperature. The conditions \eqref{Eisert} would only fail if for one of the modes the value of $b_n$ 
exceeds the sum of \emph{all} other $b_n$'s.

Let us now apply formula \eqref{Eisert} to the case of Hawking radiation. Assume first, basing on the standard black body particle-number spectrum\cite{Gray16}, that the Hawking modes are, on the average, emitted with frequencies corresponding to the peak value $\langle \omega \rangle_N(T) = (2+W(-2e^{-2})) k_\text{B} T / \hbar$, where $W$ is the Lambert $W$-function and $W(-2e^{-2}) \approx -0.4$. Alternatively, we could take the peak frequency in the energy spectrum, which gives a similar result $\langle \omega \rangle_E(T) = (3+W(-3e^{-3})) k_\text{B} T  / \hbar$ with $W(-3e^{-3}) \approx -0.18$. Consequently, we obtain $\langle b_n \rangle_N \approx 0.51$ and $\langle b_n \rangle_E \approx 0.13$. We shall denote by $\langle b \rangle$ either of these values.

If all Hawking modes are on the average typical, then condition \eqref{Eisert} is readily verified. Suppose now that there is a single `rogue' Hawking mode, the frequency $\Omega$ of which is much smaller than the expected peak value in its epoch. This is the worst-case scenario from the perspective of formula \eqref{Eisert}, because if two or more such rogue modes are emitted, their excessive values of $b_n$ will compensate each other. Formula \eqref{Eisert} then gives an explicit bound on the minimal value of $\Omega$:
\begin{align}\label{wmax}
\Omega(T)  \gtrsim \tfrac{2}{\langle b \rangle} \tfrac{k_{\text{B}} T}{\hbar} \tfrac{1}{N}
\end{align}
or, equivalently, for the maximal wavelength
\begin{align}\label{Lmax}
\Lambda(T)  \lesssim \tfrac{\pi c \hbar}{k_{\text{B}} T}  \langle b \rangle N = 8 \pi^2 \langle b \rangle N R_\text{S}(T),
\end{align}
where $R_\text{S}(T)$ is the Schwarzschild radius of the black hole in the epoch with temperature $T$. Given that the number $N$ for astrophysical black holes is 
extremely
large, condition \eqref{Lmax} is readily satisfied in every epoch of Hawking radiation. This conclusion is insensitive to the actual value of the prefactor $\langle b \rangle$. It breaks down only when the total number of emitted Hawking particles is small $N \sim 1$. For this to happen, the black hole would need to have an initial mass of the order of Planck mass $M_P$. Even in such case the radiation can still be thermal, but condition \eqref{Eisert} then imposes constraints on the frequencies of particles.

It is also instructive to compare the constraint \eqref{Eisert} with other bounds on frequencies imposed on the Planckian spectrum of Hawking radiation\cite{Visser15}. There is an insurmountable upper cut-off in the frequencies, which stems from the fact that the emitted mode cannot have an energy exceeding $Mc^2$, for a black hole of mass $M$. This translates to a minimal value of $b_n$ for each mode.

In\cite{Visser15} it is also pointed out that the Planckian spectrum of the Hawking radiation is modified by an IR cut-off in frequencies because of the ``adiabaticity constraint''. It says that the emitted Hawking particle should not feel any difference in the surface gravity $\kappa$ on the time-scale of one oscillation. More concretely, $\omega \gtrsim \omega^{\text{min}} = \sqrt{\dot{\kappa}}$, which for the Schwarzschild black hole equals to $M_P^3/M^2$.

Because $\omega^{\text{max}} \sim M$ and $\omega^{\text{min}} \sim M^{-2}$ the lower bound eventually surpasses the upper bound, which signals the ultimate breakdown of thermality. However, this happens only for Planck-scale black holes\cite{Visser15}, for which one does not expect the semi-classical approach to hold anyway.

\begin{figure}[h]
\begin{center}
\includegraphics[scale=1]{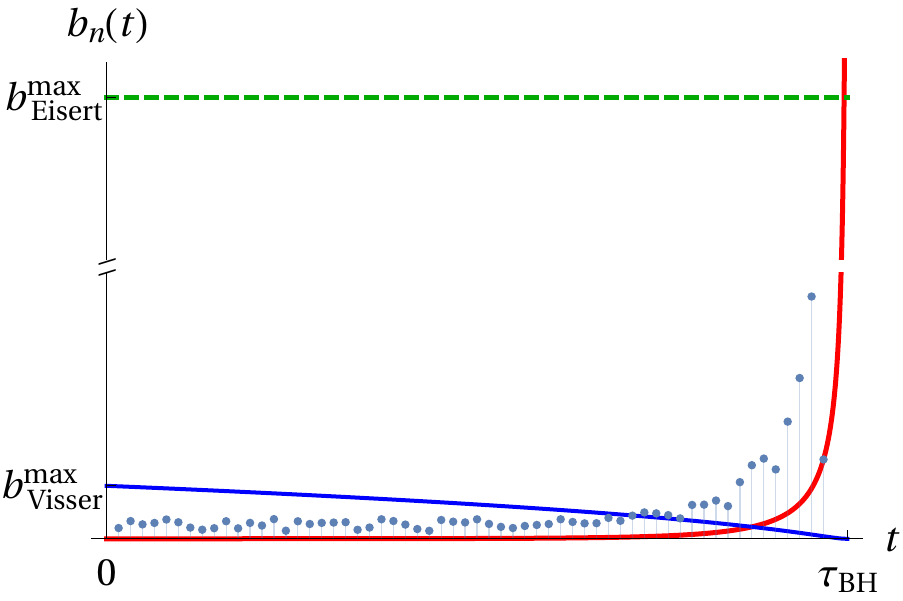}
\end{center}
\caption{\label{fig:omega}The value of the coefficients $b_n$ for the modes of the Hawking radiation is plotted against time. The value $\tau_{\text{BH}} = \frac{5120\pi G^2 M^3}{\hbar c^4}$ is the lifetime of a black hole with an initial mass $M$. The red line marks the absolute lower bound for the value of $b_n$, which corresponds to the maximal frequency. The last Hawking mode saturates this bound. The blue line depicts the adiabaticity constraint\cite{Visser15}, while the green dashed line marks the bound \eqref{Eisert}. The latter arises from the sum of all, but the largest of all $b_n$'s. When the black hole's mass is of the order of Planck mass, the adiabaticity constraint may fail, but the kinematic bound \eqref{Eisert} is still satisfied.}
\end{figure}

The bound on minimal frequency of the Hawking modes stemming from formula \eqref{Eisert} does not rely on the semi-classical approximation, hence it is more permissive than the adiabaticity constraint. In fact, it has a good chance of being satisfied even for Planck-scale microscopic black holes. A qualitative comparison between the discussed bounds is presented in Figure \ref{fig:omega}.

\pagebreak

Finally, let us point out that the analysis based on formula \eqref{Eisert} is robust against the modifications of the emission spectra due to the black hole's intrinsic spin and other physical effects, such as the grey body factors. Firstly, one can relax the condition on the global state to be a mixed Gaussian state, which induces a more general constraint\cite{Eisert2008}
\begin{align}\label{Eisert2}
\sum_{m} b_m \geq \sum_{m} \delta_m, \quad \text{ and } \quad b_n \leq \sum_{m \neq n} b_m + \delta_n - \sum_{m \neq n} \delta_m,
\end{align}
where the numbers $\delta_n \geq 0$ arise from the symplectic spectrum $\{1+\delta_1,\ldots,1+\delta_n\}$ of the global state $\ket{\Phi}$. Secondly, one could consider a more general class of non-Gaussian states, in which case constraints \eqref{Eisert2} translate to a consistency conditions between the global and local second moments of the $N$-mode quantum state.

\section{Discussion}

Our analysis is based upon two assumptions: the universal validity of quantum mechanics and the single-mode black body spectrum of the Hawking radiation. The quantum information is then preserved during the black hole formation and evaporation process, because the system's state 
follows a unitary evolution. While the first assumption, recently dubbed ``the central dogma''\cite{Maldacena20}, could and has been questioned\cite{UnruhWald17,OR}, there has been no experimental signature of the inadequacy of quantum mechanics, either at the micro-scale\cite{EPR_quarks} or in macroscopic systems\cite{Superpos2019}.
On the other hand, it is well-known that the assumption of global unitarity implies that the black body description of the Hawking radiation cannot be valid\cite{Harlow}. Nevertheless, one would expect that the deviations from the semi-classical picture should not be visible if one studies the individual modes of the Hawking radiation.

The application of the continuous-variable quantum marginal problem allowed us to rigorously demonstrate that the purity of global state of Hawking radiation is compatible with thermal reductions of all individual modes. Hence, even if the late stage of the Hawking radiation is very `improbable' from the viewpoint of the black body spectrum, it is perfectly consistent both with the global unitarity and the single-mode thermality with the corresponding Bekenstein--Hawking temperature. This is a consequence of the vast realm of pure quantum states available in $N$-mode systems with large $N$. The quantum marginal conditions are thus readily satisfied for Hawking radiation of astrophysical provenance and can accommodate highly irregular modes emitted in the last epoch. On the other hand, for black holes of microscopic origin, with masses of the order of several Planck masses, thermality induces more stringent constraints on the frequencies of individual particles.

The single-mode thermal character of the radiation does not exclude correlations between the Hawking modes. On the contrary, a typical pure $N$-mode Gaussian state will exhibit multi-mode entanglement\cite{Horodecki,GaussQI}. For instance, the shared entanglement of a single mode versus the rest of the system can be quantified with the help of the entanglement entropy\cite{GaussQI}

\begin{align}
E(b_n) = \tfrac{b_n +2}{2} \log_2 \left( \tfrac{b_n +2}{2} \right) - \tfrac{b_n}{2} \log_2 \left( \tfrac{b_n}{2} \right).
\end{align}

The method of Gaussian marginal problem applied to Hawking radiation introduced in this work opens the door to inspect the boundary between models assuming global preservation of unitary and the information loss. It is possible to assume (cf. e.g. Ref.\cite{Harlow}) that the Hawking radiation is not only single-mode thermal, but actually multi-mode thermal for a limited number modes, while being in a pure global state. To study such a hypothesis, one would have to extend the results of Ref.\cite{Eisert2008} to unveil the compatibility constraint of a global pure Gaussian state versus $k$-mode thermality, for any $1 \leq k < N$.

The solution to the multi-mode Gaussian marginal problem may offer significant advances when applied to the black hole information problem. The derivation of the maximal admissible number $k_\text{max}$ of the product thermal modes, which do not spoil the assumption of global purity, would mark the boundary of thermality versus unitarity of the Hawking radiation. It would thus provide a universal limit on the quantum-field-theoretic description, which could be tested against specific models. In other words, such a maximal number $k_\text{max}$ would indicate the maximal number of Hawking modes that one needs to measure jointly to determine whether it is consistent with global unitarity.

Although the quantum marginal problem, also in its multi-mode version, offers a kinematic constraint, it does shed some light on the admissible dynamics. As argued in Section \ref{sec:unitarity}, the global unitarity requires a departure from the LOCC paradigm and hence from the locality of the interactions. An insight into the admissible structure of correlations among the Hawking modes would offer further insight into the possible non-local dynamics of quantum information during the black hole evaporation. Because the subsequent Hawking modes should be compatible with the Bekenstein--Hawking formula for the temperature, it might also clarify the question whether the outflow of quantum information is slow and long-lasting or rather it effectively occurs only shortly before the final explosion.

One could also study the correlations in 
the final state of Hawking radiation using other tools from modern quantum optics. For instance, condition \eqref{Eisert} allows typical modes to be of similar squeezing. This may happen when the ratio $\omega_n/T_n$ does not vary much with the time step $n$. Multi-mode squeezed states constitute a well established field of quantum optics and include the so-called Gaussian graph states with the GHZ-type as a prominent example\cite{Pfister2}. From the viewpoint of the multi-mode entanglement structure, the latter exhibits highest symmetry. 
One may thus check whether a given symmetry of the final state of radiation is reflected in a specific model of unitary black hole evaporation.

Many conceptual models in the literature\cite{Giddins12a,Giddins12b,Giddins13,Avery13,TensorNet17,OsugaPage2018,Broda20} posit that the quantum degrees of freedom of a black hole are discrete, which leads to an assumption of $\Hh_{\text{BH}}$ being finite-dimensional\cite{Maldacena20}.
This suggests that realistic models involving quantised radiation may need hybrid discrete-continuous quantum information techniques\cite{Andersen2015,KunHuang19}.

\section*{Acknowledgments}

This work was supported by Swedish Research Council grant 2020-04980 (E.A.). M.E. acknowledges support of the Foundation for Polish Science under the project Team-Net NTQC number 17C1/18-00. P.H. acknowledges support by the Foundation for Polish Science (IRAP project, ICTQT, contract no. 2018/MAB/5, co-financed by EU within Smart Growth Operational Programme.

We are grateful to Don Page for his constructive comments and correspondence. E.A. thanks Bo Sundborg and Karol \.{Z}yczkowski for discussions and constructive remarks. We also thank the anonymous Reviewer for their pertinent questions and comments. 

\section*{References}

\bigskip

\bibliography{QBH}

\end{document}